\documentclass[sigconf]{acmart}
\renewcommand\footnotetextcopyrightpermission[1]{} 
\AtBeginDocument{%
  \providecommand\BibTeX{{%
    \normalfont B\kern-0.5em{\scshape i\kern-0.25em b}\kern-0.8em\TeX}}}

\setcopyright{acmcopyright}
\copyrightyear{2018}
\acmYear{2018}
\acmDOI{10.1145/1122445.1122456}

\makeatletter
\def\@copyrightspace{\relax}
\makeatother
\begin{document}

\title{SparTerm: Learning Term-based Sparse Representation for Fast Text Retrieval}


\author{Yang Bai}
\authornote{Both authors contributed equally to this research.}
\authornote{This work is done when Yang Bai is an intern at Huawei Noah’s Ark Lab.}
\affiliation{%
  \institution{Tsinghua University}
  \city{}
  \country{}
}
\email{}
\author{Xiaoguang Li}
\authornotemark[1]
\affiliation{%
  \institution{Huawei Noah’s Ark Lab}
  \city{}
  \country{}
}
\email{}
\author{Gang Wang}
\affiliation{%
  \institution{Huawei Noah’s Ark Lab}
  \city{}
  \country{}
}
\email{}
\author{Chaoliang Zhang}
\affiliation{%
  \institution{Huawei Noah’s Ark Lab}
  \city{}
  \country{}
}
\email{}
\author{Lifeng Shang}
\affiliation{%
  \institution{Huawei Noah’s Ark Lab}
  \city{}
  \country{}
}
\email{}
\author{Jun Xu}
\affiliation{%
  \institution{Renmin University of China}
  \city{}
  \country{}
}
\email{}
\author{Zhaowei Wang}
\affiliation{%
  \institution{Huawei Noah’s Ark Lab}
  \city{}
  \country{}
}
\email{}
\author{Fangshan Wang}
\affiliation{%
  \institution{Huawei Technologies Co., Ltd}
  \city{}
  \country{}
}
\email{}
\author{Qun Liu}
\affiliation{%
  \institution{Huawei Noah’s Ark Lab}
  \city{}
  \country{}
}
\email{}

\renewcommand{\shortauthors}{Trovato and Tobin, et al.}

\begin{abstract}
Term-based sparse representations dominate the first-stage text retrieval in industrial applications, due to its advantage in efficiency, interpretability, and exact term matching. In this paper, we study the problem of transferring the deep knowledge of the pre-trained language model (PLM) to \textbf{Term}-based \textbf{Spa}rse \textbf{r}epresentations, aiming to improve the representation capacity of bag-of-words(BoW) method for semantic-level matching, while still keeping its advantages. Specifically, we propose a novel framework SparTerm to directly learn sparse text representations in the full vocabulary space. The proposed SparTerm comprises an importance predictor to predict the importance for each term in the vocabulary, and a gating controller to control the term activation. These two modules cooperatively ensure the sparsity and flexibility of the final text representation, which unifies the term-weighting and expansion in the same framework. Evaluated on MSMARCO dataset, SparTerm significantly outperforms traditional sparse methods and achieves state of the art ranking performance among all the PLM-based sparse models.
\end{abstract}



\maketitle

\section{Introduction}
Text retrieval in response to a natural language query is a core task for information retrieval (IR) systems. Most recent work has adopted a two-stage pipeline to tackle this problem, where an initial set of documents are firstly retrieved from the document collection by a fast retriever, and then further re-ranked by more sophisticated models. 

For the first-stage retrieval, neural dense representations show great potentials for semantic matching and outperform sparse methods in many NLP tasks, but this is not necessarily true in scenarios that emphasize long document retrieval and exact matching\cite{Luan2020SparseDA}. Moreover, for extremely large (e.g. 10 billion) candidates collection, the dense method has to struggle with the efficiency vs. accuracy tradeoff. Classical term-based sparse representations, also known as bag-of-words (BoW), such as TF-IDF \cite{Sparckjones1972ASI} and BM25 \cite{robertson1994some}, can efficiently perform literal matching, thus playing a core role in industrial IR systems. However, traditional term-based methods are generally considered to have insufficient representation capacity and inadequate for semantic-level matching.  

\begin{figure}
  \centering
  \includegraphics[width=\linewidth]{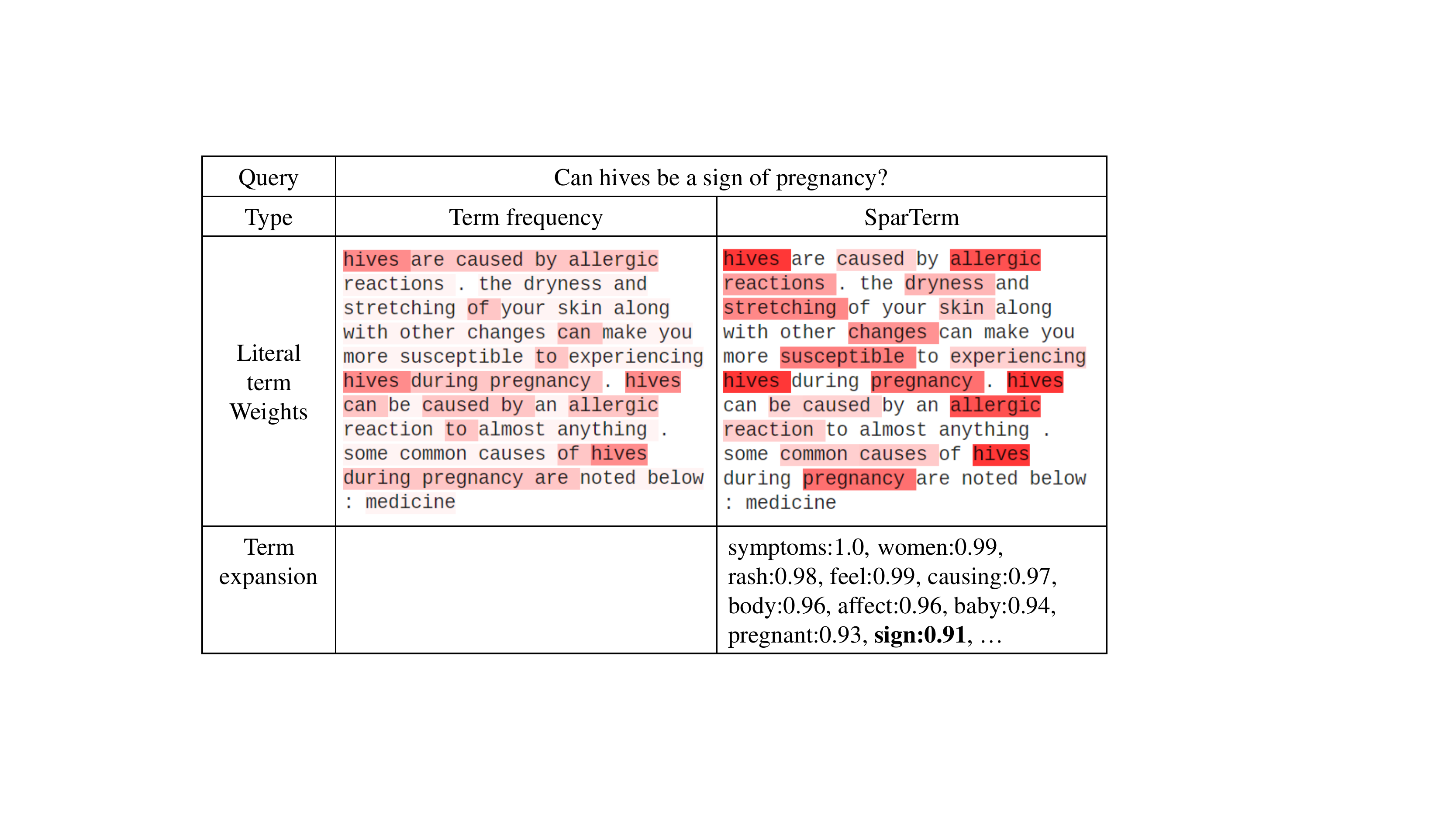}
  \caption{The comparison  between BoW and SparTerm representation. The depth of the color represents the term weights, deeper is higher. Compared with BoW, SparTerm is able to figure out the semantically important terms and expand some terms not appearing in the passage but very semantically relevant, even the terms in the target query such as ``sign''.}
  \label{fig:intro_case}
\end{figure}

Some attempts have been made to make sparse methods beyond lexical matching while still keeping their advantages. SRNM \cite{zamani2018neural} learns latent sparse representations for the query and document based on dense neural models, in which the ``latent'' token plays the role of the traditional term during inverted indexing. One challenge about SNRM is that it loses the interpretability of the original terms, which is critical to industrial systems. 

Recently proposed pre-trained language models(PLM) such as ELMO \cite{peters2018deep} and BERT \cite{devlin2018bert} show superior performance in many NLP tasks, thus providing new opportunities to transfer deep contextualized knowledge from dense representations to sparse models. Focusing on the relevant relationship between a passage/document and corresponding query, DeepCT \cite{dai2019context} and Doc2Query \cite{nogueira2019document} learn PLM-based models to enhance the performance of traditional BoW methods. The difference is that DeepCT learns a regression model to re-weight terms with contextualized representations, while Doc2query learns an encoder-decoder generative model to expand query terms for passage. Both of these two methods train an auxiliary intermediate model and then help refine the final sparse representations to achieve better text ranking performance.

In this paper, we propose a novel framework SparTerm to learn \textbf{Term}-based \textbf{Spa}rse \textbf{r}epresentations directly in the full vocabulary space. Equipped with the pre-trained language model, the proposed SparTerm learns a function to map the frequency-based BoW representation to a sparse term importance distribution in the whole vocabulary, which offers the flexibility to involve both term-weighting and expansion in the same framework. As shown in Figure ~\ref{fig:intro_case}, compared with BoW representation, SparTerm assigns more weights to the term of high distinguishability given the context,  and expand extra terms hopefully bridging the lexical gap with future queries. We empirically show that SparTerm significantly increase the upper limit of sparse retrieval methods, and gives new insights of transferring deep knowledge from PLM-based representation to simple BoW representations.

More specifically, SparTerm comprises an importance predictor and a gating controller. The importance predictor maps the raw input text to a dense importance distribution in the vocabulary space, which is different from traditional term weighting methods that only consider literal terms of the input text. To ensure the sparsity and flexibility of the final representation, the gating controller is introduced to generate a binary and sparse gating signal across the dimension of vocabulary size, indicating which tokens should be activated. These two modules cooperatively yield a term-based sparse representation based on the semantic relationship of the input text with each term in the vocabulary.

\textbf{Our contributions.} In summary, we propose to directly learn term-based sparse representation in the full vocabulary space. The proposed SparTerm indicates that there is much space for improving the ranking performance of termed-based representations, while still keeping the interpretability and efficiency of BoW methods. Evaluated on MSMARCO \cite{NguyenRSGTMD16} dataset, SparTerm significantly outperforms previous sparse models based on the comparable size of PLMs. The top-ranking performance of SparTerm even outperforms Doc2Query-T5, which is based on the pre-trained model of 2x model size and 70x pre-training corpus size. Moreover, we conduct further empirical analysis about how the deep knowledge of PLMs can be transferred to the sparse method, which gives new insights for sparse representation learning.  





\section{Related Work}

Our work relates to two research fields: bag-of-words representations and pre-trained language model for text retrieval. 
\subsection{Bag-of-words Methods}
Bag-of-words(BoW) methods have played a central role in the first-stage retrieval. These methods convert a document or query into a set of single terms, and each term associates a weight to characterize its weight. Most of the early common practice adopted TF-IDF style models to calculate weights. Robertson \cite{robertson1994some} proposed the well-known method BM25, which further improve the performance of the original TF-IDF. Later proposed methods, such as \cite{lafferty2001document}, \cite{zaragoza2004microsoft}, \cite{strohman2005indri}, did not show much advantage over BM25. More recently, Hamed Zamani \cite{zamani2018neural} proposed SRNM to learn a sparse coding in hidden space using weak supervision, which shows good potential for solving the ``lexical mismatch'' problem. However, the latent unexplainable tokens can not ensure that documents with exact matched terms can be retrieved.

\subsection{PLMs for dense text retrieval}
The pre-trained language models like BERT \cite{devlin2018bert} show new possibilities for text retrieval. Based on dense representations, Lee \cite{lee2019latent} proposed ORQA with bi-encoder architecture to retrieve candidate passages for question answering using FAISS \cite{JDH17}. However, analysis from \cite{Luan2020SparseDA} concludes that bi-encoders based on dense representation suffer from its capacity limitation in scenarios that emphasize long document retrieval and exact matching. Following the late-interaction paradigm, Khattab \cite{Khattab2020ColBERTEA} proposed Col-BERT to conduct efficient interaction between the query and document, which can run 150x faster than fully-interactive BERT but achieve comparable precision. Though much faster than BERT, Col-BERT is still not computationally feasible for large scale first-stage retrieval, for the existence of the late interaction layer.

\begin{figure*}[h]
  \centering
  \includegraphics[scale=0.6]{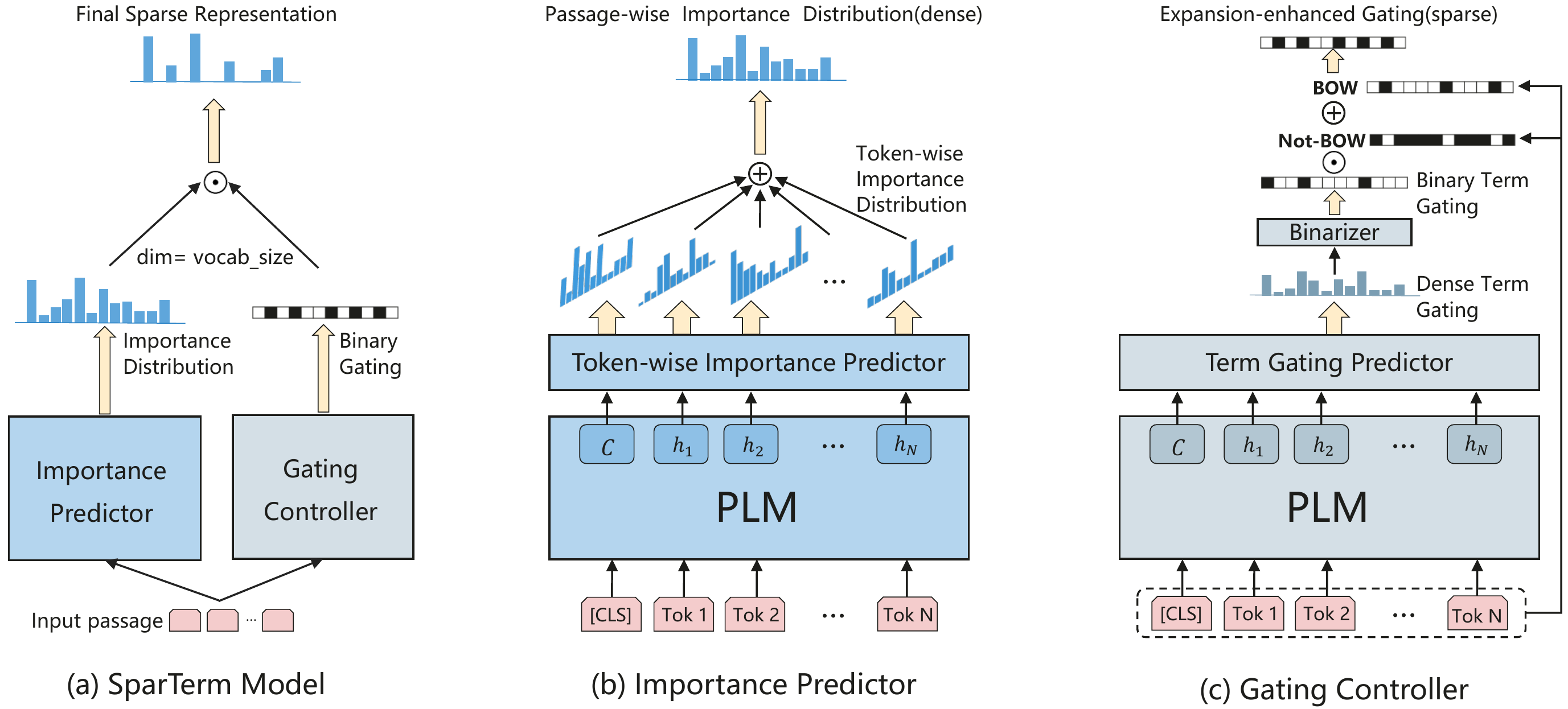}
  \caption{Model Architecture of SparTerm. Our overall architecture contains an importance predictor and a gating controller. The importance predictor generates a dense importance distribution with the dimension of vocabulary size, while the gating controller outputs a sparse and binary gating vector to control term activation for the final representation. These two modules cooperatively ensure the sparsity and flexibility of the final representation.}
  \Description{Model Architecture of SparTerm}
  \label{fig:overview}
\end{figure*}

\subsection{PLMs for sparse text retrieval}
Several PLM-based models have emerged to improve the traditional sparse BoW representations. Dai \cite{dai2019context} proposed DeepCT to estimate a term's weight considering its contextualized information, and this work was later extended to generate document-level term weights \cite{dai2020context}. Another work Doc2query \cite{nogueira2019document} tries to ``translate'' potential queries to expand document content, which also shows a large improvement compared to the traditional BM25 method. The biggest difference between our work and these two methods is that DeepCT and Doc2Query train an auxiliary intermediate model to help refine the sparse representations, while SparTerm is desinged to directly learn sparse representations within the whole vocabulary.

\section{Sparse Representation Learning}
This section presents the model architecture of SparTerm and the corresponding training strategy.
\subsection{Overview}
Figure~\ref{fig:overview}(a) depicts the general architecture of SparTerm which comprises an importance predictor and a gating controller. Given the original textual passage $p$, we aim to map it into a deep and contextualized sparse representation $p{'}$ in the vocabulary space. The mapping process can be formulated as:
\begin{equation}
p{'} = \mathcal{F}(p) \odot \mathcal{G}(p)
\label{eq:task_definition}
\end{equation}
where $\mathcal{F}$ is the item importance predictor and $\mathcal{G}$ the gating controller. The importance predictor $\mathcal{F}$ generates a dense vector representing the semantic importance of each item in the vocabulary. The gating controller $\mathcal{G}$ generates a binary gating vector to control which terms to appear in the final sparse representation. To achieve this, we let $||\mathcal{G}(p)||<\lambda$ and $\mathcal{G}(p)\in \{0,1\}^{v}$, where $\lambda$ is the maximum number of non-zero elements for $p{'}$, and $v$ the vocabulary size. These two modules cooperatively ensure the sparsity and flexibility of the final representation $p{'}$. We discuss the detailed model architecture and learning strategy for $\mathcal{F}$ and $\mathcal{G}$ in the following sections.


\begin{table*}

\begin{tabular}{|l|l|}
\hline
\textbf{Term expansion} & \textbf{Description and examples} \\ \hline
Passage2query & Expand words that tend to appear in corresponding queries, i.e. ``how far'', ``what causes''. \\ \hline
Synonym & Expand synonym for original core words, i.e. ``cartoon''->``animation''. \\ \hline
Co-occurred words & Expand frequently co-occurred words for original core words, i.e. ``earthquakes''->``ruins''. \\ \hline
Summarization words & Expand summarization words that tend to appear in passage summarization or taggings. \\ \hline
\end{tabular}
\caption{Different kinds of term expansion.}
\label{tab:expansion}
\end{table*}

\subsection{The Importance Predictor}
Given the input passage $p$, the importance predictor outputs semantic importance of all the terms in the vocabulary, which unify term weighting and expansion into the framework. As shown in Figure ~\ref{fig:overview}(b), prior to importance prediction, BERT-based encoder is employed to help get the deep contextualized embedding $h_{i}$ for each term $w_{i}$ in the passage $p$. Each $h_{i}$ models the surrounding context from a certain position $i$, thus providing a different view of which terms are semantically related to the topic of the current passage. With a token-wise importance predictor, we obtain a dense importance distribution $I_{i}$ of dimension $v$ for each $h_{i}$:
\begin{equation}
I_{i}=Transform(h_{i})E^\mathrm{T}+b
\label{eq:token-wise distri}
\end{equation}  
where $Transform$ denotes a linear transformation with GELU activation and layer normalization, $E$ is the shared word embedding matrix and $b$ the bias term. Note that the token-wise importance prediction module is similar to the masked language prediction layer in BERT, thus we can initialize this part of parameters directly from pre-trained BERT. The final passage-wise importance distribution can be fetched simply by the summation of all token-wise importance distributions:
\begin{equation}
I=\sum_{i=0}^{L}Relu(I_{i})
\label{eq:passage-wise distri}
\end{equation} 
where $L$ is the sequence length of passage $p$ and Relu activation function is leveraged to ensure the nonnegativity of importance logits. 
\subsection{The Gating Controller}
The gating controller generates a binary gating signal of which terms to activate to represent the passage. 
First, the terms appearing in the original passage, which we referred to as literal terms, should be activated by the controller by default. 
Apart from the literal terms, some other terms related to the passage topic are also expected to be activated to tackle the ``lexical mismatch'' problem of BOW representation. Accordingly, we propose two kinds of gating controller: literal-only gating and expansion-enhanced gating, which can be applied in scenarios with different requirements for lexical matching. 

\textbf{Literal-only Gating.} If simply setting $\mathcal{G}(p)=BOW(p)$, where $BoW(p)$ denotes the binary BoW vector for passage $p$, we get the literal-only gating controller. In this setting, only those terms existing in the original passage are considered activated for the passage representation. Without expansion for non-literal terms, the sparse representation learning is reduced to a pure term re-weighting scheme. Nevertheless, in the experiment part, we empirically show that this gating controller can achieve competitive retrieval performance by learning importance for literal terms.

\textbf{Exapnsion-enhanced Gating.} The expansion-enhanced gating controller activates terms that can hopefully bridge the ``lexical mismatch'' gap. Similar to the importance prediction process formulated by Equation \eqref{eq:token-wise distri} and Equation \eqref{eq:passage-wise distri}, we obtain a passage-wise dense term gating distribution $G$ of dimension $v$ with independent network parameters, as shown in Figure~\ref{fig:overview}(c). Note that although the gating distribution $G$ and the importance distribution $I$ share the same dimension $v$, they are different in logit scales and mathematical implications. $I$ represents the semantic importance of each term in vocabulary, while $G$ quantifies the probability of each term to participate in the final sparse representation. To ensure the sparsity of $p^{'}$, we apply a binarizer to $G$:
\begin{equation}
G^{'}=Binarizer(G)
\label{eq:sparse G}
\end{equation} 
where the $Binarizer$ denotes a binary activation function which outputs only 0 or 1. The gating vector for expansion terms $G_{e}$ is obtained by:
\begin{equation}
G_{e}=G^{'} \odot (\neg BoW(p))
\label{eq:final G}
\end{equation}
where the bitwise negation vector $\neg BoW(p)$ is applied to ensure orthogonal to the literal-only gating. Simply adding the expansion gating and the literal-only gating, we get the final expansion-enhanced gating vector $G_{le}$:
\begin{equation}
G_{le}=G_{e}+ BoW(p)
\label{eq:final G}
\end{equation}
Involving both literal and expansion terms, the final sparse representation can be a ``free'' distribution in the vocabulary space. Note that in the framework of SparTerm, expanded terms are not directly appended to the original passage, but are used to control the gating signal of whether allowing a term participating the final representation. This ensures the input text to the BERT encoder is always the natural language of the original passage.
\subsection{Training}
In this section, we introduce the training strategy of the importance predictor and expansion-enhanced gating controller.

\textbf{Training the importance predictor.} The importance predictor is trained end-to-end by optimizing the ranking objective. Let $ R=\{(q_{1},p_{1,+},p_{1,-}), ...,(q_{N},p_{N,+},p_{N,-})\} $ denote a set of N training instances; each containing a query $q_{i}$, a posotive candidate passage $p_{i,+}$ and a negative one $p_{i,-}$, indicating that $p_{i,+}$ is more relevant to the query than $p_{i,-}$. The loss function is the negative log likelihood of the positive passage:
\begin{equation}
L_{rank}(q_{i},p_{i,+},p_{i,-})= -\log \frac{e^{sim(q_{i}^{'},p_{i,+}^{'})}}{e^{sim(q_{i}^{'},p_{i,+}^{'})}+e^{sim(q_{i}^{'},p_{i,-}^{'})}}
\label{eq:loss_rank}
\end{equation}
where $q_{i}^{'}$, $p_{i,+}^{'}$, $p_{i,-}^{'}$ is the sparse representation of $q_{i}$, $p_{i,+}$, $p_{i,-}$ obtained by Equation \eqref{eq:task_definition}, $sim$ denotes any similarity measurement such as dot-product. Different with the training objective of DeepCT \ref{dai2019context}, we don't directly fit the statistical term importance distribution, but view the importance as intermediate variables that can be learned by distant supervisory signal for passage ranking. End-to-end learning can involve every terms in the optimization process, which can yield smoother importance distribution, but also of enough distinguishability.   

\textbf{Training the exapnsion-enhanced gating controller.} We summarize four types of term expansion in Table \ref{tab:expansion}, all of which can be optimized in our SparTerm framework. Intuitively, the pre-trained BERT already has the ability of expanding synonym words and co-occured words by the Masked Language Model(MLM) pre-training task. Therefore, in this paper, we focus on expanding passage2query-alike and summarization terms. Given a passage-query/summary parallel corpus $\mathbf{C}$, where $p$ is a passage, $t$ the corresponding target text, and $T$ of dimension $v$ is the binary bag-of-words vector of $t$. We use the binary cross-entropy loss to maximize probability values of all the terms in vocabulary:
\begin{equation}
L_{exp}=-\lambda_{1}\sum\nolimits_{j\in \{m|T_{m}=0\}}log(1-G_{j})-  \lambda_{2}\sum\nolimits_{k\in \{m|T_{m}=1\}}logG_{k} 
\label{eq:loss_exp}
\end{equation} 
where $G$ is the dense gating probability distribution for $p$, $\lambda_{1}$ and $\lambda_{2}$ two tunable hyper-parameters. $\lambda_{1}$ is the loss weight for terms expected not to be expanded, while $\lambda_{2}$ is for terms that appear in the target text. In the experiment, we set $\lambda_{2}$ much larger than $\lambda_{1}$ to encourage more terms to be expanded. 

\textbf{End-to-end joint training.} Intuitively, the supervisory ranking signal can also be leveraged to guide the training of the gating controller, thus we can train the importance predictor and gating controller jointly:
\begin{equation}
L=L_{rank}+L_{exp}
\label{eq:loss}
\end{equation}

\begin{table*}[t]
	\centering
	\begin{tabular}{lcccccccc}
		\hline
		Model                       & MRR@10 & R@10  & R@20  & R@50  & R@100 & R@200 & R@500 & R@1000 \\ \hline
		BM25                        & 18.6   & -     & 49    & 60    & 69    & 75    & 82    & 85.71  \\
		Doc2query                   & 21.5   & -     & -     & -     & -     & -     & -     & 89.1   \\
		Doc2query-T5                & 27.7   & -     & -     & 75.6  & 81.89 & 86.88 & 91.64 & 94.7   \\
		DeepCT                      & 24.3   & 49    & 58    & 69    & 76    & 82    & 86    & 91     \\ \hline
		SparTerm(literal-only)       & 27.46  & 51.05 & 60.21 & 71.55 & 78.28 & 83.27 & 88.33 & 91.16  \\
		SparTerm(expansion-only)     & 19.8   & 40.93 & -     & 63.42 & 70.96 & 77.62 & 84.81 & 89.08  \\
		SparTerm(expansion-enhanced) & \textbf{27.94}  & 51.95 & 61.58 & 72.48 & 78.95 & 84.05 & 89.5  & 92.45  \\ \hline
	\end{tabular}
	\caption{\label{overall}Performances of different models on Dev Set of MSMARCO Passage Retrieval dataset.}
\end{table*}

\section{Experimental Setup}
\subsection{Datasets and Metrics}
We evaluate our method on \textbf{MSMARCO} \cite{NguyenRSGTMD16} which consists of two benchmark datasets:

\textbf{MSMARCO Passage Retrieval dataset} is based on the public MSMARCO dataset with a collection of 8.8M passages from Web pages gathered from Bing’s results to 1M real-world queries. Each query is associated with one or very few passages marked as relevant while no passage explicitly indicated as irrelevant. We build a small dev set for validating the full ranking performance instead of re-ranking by sampling the most relevant 1M passages to 1000 queries from the original passage ranking dev set with BM25.

\textbf{MSMARCO Document Retrieval dataset} is based on the source documents which contain the passages in the passage retrieval task. The dataset contains 367,013 documents and 367,013 queries for training set and 5,193 queries for dev set.

The original Dev Set of MSMARCO dataset is a re-ranking task, which is inconsistent with the retrieval task. Therefore, to find the best checkpoint of our model more accurately we build a new Dev Set to evaluate the retrieval performance by sampling about 1M passages from the collections and 1,000 queries from the original Dev Set (including the top 1000 passages of each query retrieved by BM25).
To evaluate the full ranking performance of our model, we use the sparse representation of each document to build the inverted index and use the sparse representation of queries to retrieval topK relevant documents and measure the performance with MRR@10 and Recall from top10 to top1000.

\subsection{Implementation}
The Importance Predictor and Gating Controller of our model have the same architecture and hyper-parameters of BERT (12-layer, 768-hidden, 12-heads, 110M parameters) and do not share weights. We initialize the Importance Predictor with Google’s official pre-trained $\mathrm{BERT_{base}}$ model while the parameters of Token-wise Importance Predictor are initialized with the Masked Language Prediction layer of BERT. 
When using expansion-enhanced gating, the Gating Controller is also initialized with $\mathrm{BERT_{base}}$. 
We fine-tune our model on the training set of MSMARCO passage retrieval dataset on 4 NVIDIA-v100 GPUs with a batch size of 128.
During the fine-tuning, we first fine-tune the Gating Controller with Equation \eqref{eq:loss_exp} for 50k iterations where $\lambda_{1}=1e-3$ and $\lambda_{2}=1$. Then we fix the parameters of the Gating Controller and fine-tune our SparTerm jointly for 100k iterations. 
We use Adam optimizer with the learning rate $2\times10^{-5}$. To ensure the sparsity, the threshold in the Binarizer in Equation \eqref{eq:sparse G} is set to $0.7$.
We do not fine-tune our model on the training set of document retrieval dataset but just use the model trained on the passage retrieval dataset for the document ranking.


\subsection{Baselines and Experimental Settings}
We compare our model with the following strong baselines which are all methods based on sparse representation . The former two focus on re-weighting while the latter two focus on document expansion:

\begin{itemize}
\item \textbf{BM25}\cite{robertson1994some} is a bag-of-words retrieval models with frequency-based signals to estimate the weights of terms in a text.

\item \textbf{DeepCT}\cite{dai2019context} is a deep contextualized term weighting model which maps the BERT's representations to term weightings for retrieval.

\item \textbf{Doc2query}\cite{nogueira2019document} is a document expansion method with Transformer that can expand documents with terms related to the documents' content.

\item \textbf{Doc2query-T5}\cite{Cheriton2019FromDT} is a document expansion method which utilizes more powerful T5 \cite{Raffel2019ExploringTL} language model to generate queries for document expansion.
\end{itemize}

We also evaluate three different settings of SparTerm for evaluation:
\begin{itemize}
	\item \textbf{SparTerm(literal-only)} uses Importance Predictor with the Literal-only Gating which can also be seen as a term weighting model.
	
	\item \textbf{SparTerm(expansion-only)} uses the Expansion-enhanced Gating for passage expansion without term weighting. We just add the expanded words (weight of each word is 1) to the passages.
	
	\item \textbf{SparTerm(expansion-enhanced)} implements both Importance Predictor and Expansion-enhanced Gating for sparse representation of passage.
\end{itemize}

\section{Experimental Results}
\subsection{Performance on Passage Full Ranking}
Table \ref{overall} shows the full ranking performances of our models and baselines on MSMARCO Passage Retrieval dataset.
SparTerm (expansion-enhanced) outperforms all baselines on MRR, achieving the state-of-the-art ranking performance among all sparse models, and outperforms all baselines except Doc2query-T5 on Recall. 
We find that SparTerm achieves more significant performance improvements on MRR and Recall@10-100, which illustrates that our model has a more significant ability on top ranking compared with previous sparse models.
Further, pre-trained language model(PLM) based methods (DeepCT, Doc2query-T5, and SparTerm) perform better than those without PLM, demonstrating that PLM can facilitate the passage full ranking with better representation. Considering the improvements T5 brings to Doc2query, we believe that SparTerm can be further improved with more advanced PLM.

Even without any expansion, SparTerm(literal-only) outperforms DeepCT on both MRR and Recall, demonstrating that SparTerm can produce more effective term weights thus facilitating the retrieval. We also analyze the difference between SparTerm and DeepCT on term weighting in Section \ref{weighting}.
With only the expanded words, SparTerm achieves a definite improvement compared with BM25, especially on Recall. This improvement proves the effectiveness of passage expansion on improving the Recall for retrieval.


\begin{table}[]
	\begin{tabular}{l|c}
		\hline
		Model                                                  & MRR@10 \\ \hline
		BM25+\textit{PassageRetrievalMax}                      & 23.6   \\
		HDCT+\textit{PassageRetrievalMax}                      & 26.1   \\
		BM25                                                   & 24.5   \\
		HDCT(sum)                                              & 28.0   \\
		HDCT(decay)                                            & 28.7   \\ \hline
		SparTerm(literal-only)+\textit{PassageRetrievalMax}       & 28.5  \\
		SparTerm(expansion-enhanced)+\textit{PassageRetrievalMax} & \textbf{29.0}  \\ \hline
	\end{tabular}
\caption{\label{docretrieval} Performance of baselines and our models on dev set of MSMARCO document ranking dataset. All use the max score of passages in the document as the document score at the query time.}
\end{table}

\subsection{Performance on Document Ranking}
For the Document Ranking task, we cut down each document into several passages to adapt the max length (256) of the sequence of our model and generate the sparse representation of each passage with our model. We compare our models with two baseline methods: BM25 \cite{robertson1994some} and HDCT \cite{dai2020context}. HDCT is based on the work of DeepCT and focuses on document ranking, which is also a term weighting method. HDCT compares two different ways to combine the representations of passages for document ranking. The first one represents the document as a sum of the passage representations while the second one uses a decayed weighted sum. The \textit{PassageRetrievalMax} does not represent the document but just calculates the scores of passages in the document and choose the maximum score as the score of the document for ranking. Table \ref{docretrieval} shows the ranking performance of baselines and our models. Here we only report the results of \textit{PassageRetrievalMax} of our models.

Strictly speaking, it is incomparable between HDCT and our models since we fine-tune SparTerm on MSMARCO pasage ranking dataset while HDCT was trained using document titles on MARCO. Even though, SparTerm(expansion-enhanced) still achieves a better performance on document ranking compared with HDCT, demonstrating that the sparse representation produced by SparTerm can also facilitate long document retrieval.

\begin{table}[]
	\begin{tabular}{l|cc}
		\hline
		Model                    & MRR@10 & R@1000 \\ \hline
		Query-tf                 & 25.7   & 94.2   \\
		Query-neural-symmetric   & \textbf{26.4} & \textbf{94.7} \\
		Query-neural-asymmetric  & 25.4   & 94.2  \\ \hline
	\end{tabular}
	\caption{\label{query_representation}Performances of our model with different query representation strategies on our new Dev Set of MSMARCO passage retrieval.}
\end{table}


\begin{figure*}[h]
	\centering
	\includegraphics[scale=0.4]{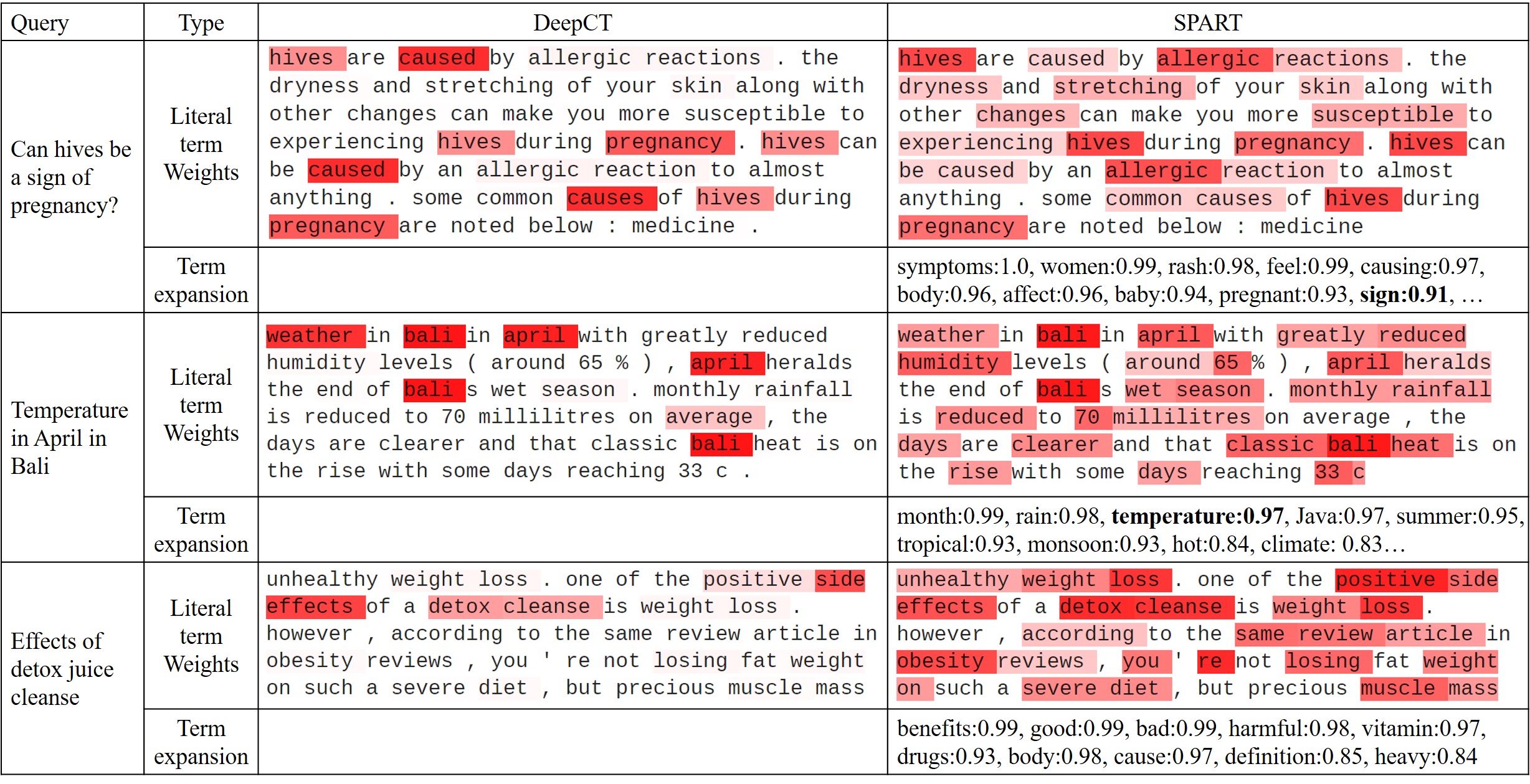}
	\caption{Term weightings of different passages weighted by DeepCT and SparTerm, and the expanded terms with their probabilities (before the binarization) predicted by SparTerm. The depth of the color represents the term weights, deeper is higher.}
	\label{heat}
\end{figure*}


\subsection{Comparison of Different Query Representation Methods}
We conduct experiments to evaluate the performance of SparTerm with different query representation methods:
\begin{itemize}
	\item \textbf{Query-tf} is a one-tower model that use tf-based vectors to represent the queries while use the model to represent documents.
	
	\item \textbf{Query-neural-symmetric} is a symmetric two-tower model to represent queries and passages that the two towers with the same architectures share the same weights.
	
	\item \textbf{Query-neural-asymmetric} is a asymmetric two-tower model that the two towers do not share weights. Queries and passages are represented with different towers.
\end{itemize}

The results are reported in Table \ref{query_representation}, from where we find that the neural representation of queries with the symmetric two-tower model brings better performance on MRR and Recall on our built Dev set. The symmetric model performs better than the asymmetric model might because asymmetric two-tower architecture leads to twice the quantity of parameters, which makes the model more difficult to converge.
We further analyze the distribution of passage term weights with different query representation methods and find that tf-based representation of query results in a sharper distribution compared to the neural representation.  The reason may be that the query representation is fixed during training, the model needs to give more weights to the relative terms in the positive passage.

\subsection{Analysis of Term Weighting} \label{weighting}
To further evaluate the ability of SparTerm on term weighting, we normalize the term weights of passages weighted by DeepCT and SparTerm(literal-only) to the same range and visulization them in Figure \ref{heat}. Figure \ref{heat} shows three different queries(the first column) and the most relevant passages. The depth of the color represents the weights of terms, deeper is higher. We find that both DeepCT and SparTerm can figure out the most important terms and give them higher weights. However, DeepCT obtains sparser and sharper distributions and only activates very few terms in a passage, missing some important terms, such as ``allergic reaction'' in the first case. SparTerm can yield a smoother importance distribution by activating more terms though not appearing in the query. This distribution allows the passage to be retrieved by more queries. This also demonstrates that our model has a better ability on pointing out important terms in a passage.

\begin{figure*}[h]
	\centering
	\includegraphics[scale=0.52]{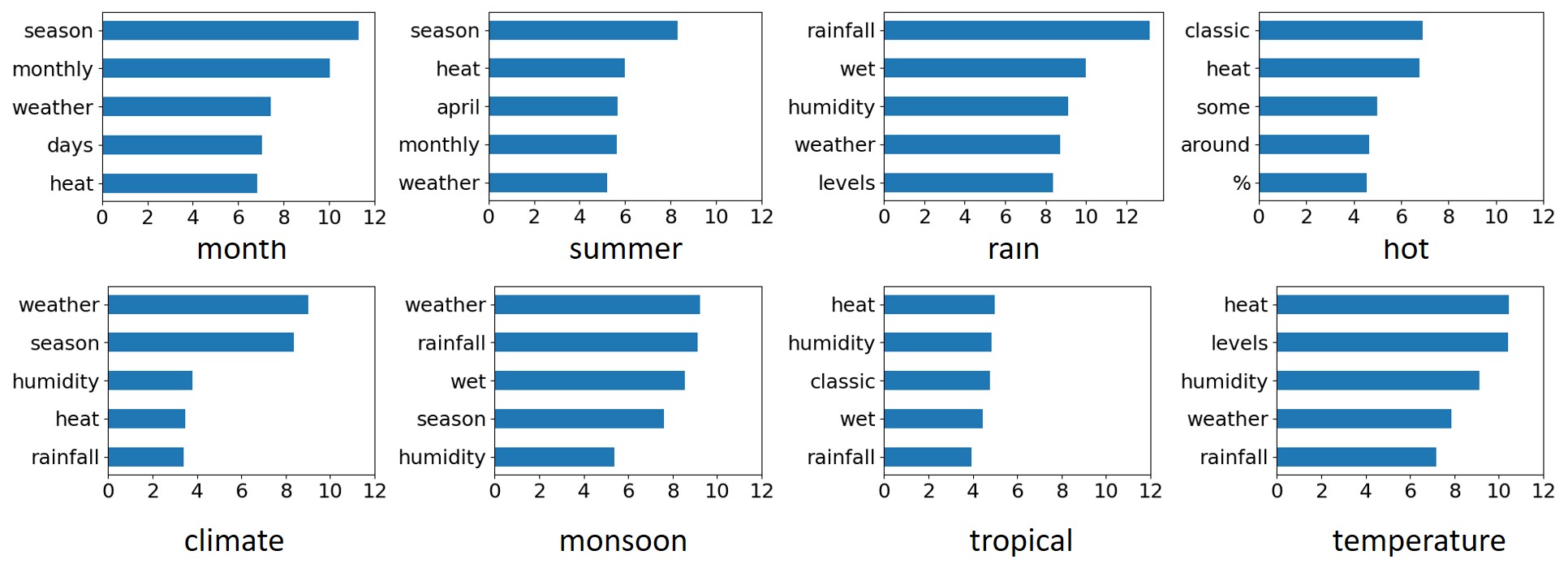}
	\caption{The Top 5 contributing words to the expanded words of the second case in Figure \ref{heat}. The X-axis are the words in the passage and Y-axis represents logit.}
	\label{contribution}
\end{figure*}

\subsection{Analysis of Term Expansion}
Figure \ref{heat} shows the expanded terms and their probabilities for different passages predicted by the Gating Controller. The probability of each term illustrates how likely this term to be expanded. It is obvious that our model can really activate some important terms not appearing in the passage but very semantically similar, especially occurring in the queries such as ``sign'' in the first case and ``temperature'' in the second case.

In order to analyze how these words are expanded and which category in Figure \ref{heat} do they belong to, we trace the source of each expanded word and show the top 5 words with their logits which contribute to the expanded word in Figure \ref{contribution}. 
We can find that there are basically three different situations of the expanded terms: 

\textbf{(1)} The passage2query terms such as ``temperature'': Almost every word in the passage contributes much to this kind of terms, which seem more likely to learn from the supervised signal.

\textbf{(2)} Synonyms of the original terms, i.e. ``weather'' and ``climate'', ``rainfall'' and ``rain'', ``season, monthly'' and ``month'', ``heat'' and ``hot''.

\textbf{(3)} Co-occurred words for the original terms, i.e. ``season, heat''->``summer'', ``wet, humidity, weather''->``rain'' and ``heat, rainfall, humidity''->``tropical, monsoon''. 

The first situation is benefited by the optimization objective of the Gating Controller while the latter two are more likely the ability of MLM pretraining task since we reuse the MLM module for prediction in the Gating Controller.

\section{Conclusion}
In this work, we propose SparTerm to directly learn term-based sparse representation in the full vocabulary space. SparTerm learns a function to map the frequency-based and BoW representation to a sparse term importance distribution in the whole vocabulary space, which involves both term-weighting and expansion in the same framework.
Experiments conducted on MSMARCO dataset show that SparTerm significantly outperforms previous sparse models based on the comparable size of PLMs, achieving state-of-the-art ranking performance among all sparse models. We conduct further empirical analysis about how the deep knowledge of PLMs can be transferred to the sparse method, which gives new insights for sparse representation learning.
Empirical results show that SAPRT significantly increases the upper limit of sparse retrieval methods.

\section*{Acknowledgement}
We thank Xin Jiang, Xiuqiang He, and Xiao Chen for the helpful discussions.

\bibliographystyle{ACM-Reference-Format}
\bibliography{SparTerm}


\begin{thebibliography}{18}


\ifx \showCODEN    \undefined \def \showCODEN     #1{\unskip}     \fi
\ifx \showDOI      \undefined \def \showDOI       #1{#1}\fi
\ifx \showISBNx    \undefined \def \showISBNx     #1{\unskip}     \fi
\ifx \showISBNxiii \undefined \def \showISBNxiii  #1{\unskip}     \fi
\ifx \showISSN     \undefined \def \showISSN      #1{\unskip}     \fi
\ifx \showLCCN     \undefined \def \showLCCN      #1{\unskip}     \fi
\ifx \shownote     \undefined \def \shownote      #1{#1}          \fi
\ifx \showarticletitle \undefined \def \showarticletitle #1{#1}   \fi
\ifx \showURL      \undefined \def \showURL       {\relax}        \fi
\providecommand\bibfield[2]{#2}
\providecommand\bibinfo[2]{#2}
\providecommand\natexlab[1]{#1}
\providecommand\showeprint[2][]{arXiv:#2}

\bibitem[\protect\citeauthoryear{Cheriton}{Cheriton}{2019}]%
        {Cheriton2019FromDT}
\bibfield{author}{\bibinfo{person}{D. Cheriton}.}
  \bibinfo{year}{2019}\natexlab{}.
\newblock \showarticletitle{From doc2query to docTTTTTquery}.
\newblock


\bibitem[\protect\citeauthoryear{Dai and Callan}{Dai and Callan}{2019}]%
        {dai2019context}
\bibfield{author}{\bibinfo{person}{Zhuyun Dai} {and} \bibinfo{person}{Jamie
  Callan}.} \bibinfo{year}{2019}\natexlab{}.
\newblock \showarticletitle{Context-aware sentence/passage term importance
  estimation for first stage retrieval}.
\newblock \bibinfo{journal}{\emph{arXiv preprint arXiv:1910.10687}}
  (\bibinfo{year}{2019}).
\newblock


\bibitem[\protect\citeauthoryear{Dai and Callan}{Dai and Callan}{2020}]%
        {dai2020context}
\bibfield{author}{\bibinfo{person}{Zhuyun Dai} {and} \bibinfo{person}{Jamie
  Callan}.} \bibinfo{year}{2020}\natexlab{}.
\newblock \showarticletitle{Context-Aware Document Term Weighting for Ad-Hoc
  Search}. In \bibinfo{booktitle}{\emph{Proceedings of The Web Conference
  2020}}. \bibinfo{pages}{1897--1907}.
\newblock


\bibitem[\protect\citeauthoryear{Devlin, Chang, Lee, and Toutanova}{Devlin
  et~al\mbox{.}}{2018}]%
        {devlin2018bert}
\bibfield{author}{\bibinfo{person}{Jacob Devlin}, \bibinfo{person}{Ming-Wei
  Chang}, \bibinfo{person}{Kenton Lee}, {and} \bibinfo{person}{Kristina
  Toutanova}.} \bibinfo{year}{2018}\natexlab{}.
\newblock \showarticletitle{Bert: Pre-training of deep bidirectional
  transformers for language understanding}.
\newblock \bibinfo{journal}{\emph{arXiv preprint arXiv:1810.04805}}
  (\bibinfo{year}{2018}).
\newblock


\bibitem[\protect\citeauthoryear{Johnson, Douze, and J{\'e}gou}{Johnson
  et~al\mbox{.}}{2017}]%
        {JDH17}
\bibfield{author}{\bibinfo{person}{Jeff Johnson}, \bibinfo{person}{Matthijs
  Douze}, {and} \bibinfo{person}{Herv{\'e} J{\'e}gou}.}
  \bibinfo{year}{2017}\natexlab{}.
\newblock \showarticletitle{Billion-scale similarity search with GPUs}.
\newblock \bibinfo{journal}{\emph{arXiv preprint arXiv:1702.08734}}
  (\bibinfo{year}{2017}).
\newblock


\bibitem[\protect\citeauthoryear{Khattab and Zaharia}{Khattab and
  Zaharia}{2020}]%
        {Khattab2020ColBERTEA}
\bibfield{author}{\bibinfo{person}{O. Khattab} {and} \bibinfo{person}{M.
  Zaharia}.} \bibinfo{year}{2020}\natexlab{}.
\newblock \showarticletitle{ColBERT: Efficient and Effective Passage Search via
  Contextualized Late Interaction over BERT}.
\newblock \bibinfo{journal}{\emph{Proceedings of the 43rd International ACM
  SIGIR Conference on Research and Development in Information Retrieval}}
  (\bibinfo{year}{2020}).
\newblock


\bibitem[\protect\citeauthoryear{Lafferty and Zhai}{Lafferty and Zhai}{2001}]%
        {lafferty2001document}
\bibfield{author}{\bibinfo{person}{John Lafferty} {and}
  \bibinfo{person}{Chengxiang Zhai}.} \bibinfo{year}{2001}\natexlab{}.
\newblock \showarticletitle{Document language models, query models, and risk
  minimization for information retrieval}. In
  \bibinfo{booktitle}{\emph{Proceedings of the 24th annual international ACM
  SIGIR conference on Research and development in information retrieval}}.
  \bibinfo{pages}{111--119}.
\newblock


\bibitem[\protect\citeauthoryear{Lee, Chang, and Toutanova}{Lee
  et~al\mbox{.}}{2019}]%
        {lee2019latent}
\bibfield{author}{\bibinfo{person}{Kenton Lee}, \bibinfo{person}{Ming-Wei
  Chang}, {and} \bibinfo{person}{Kristina Toutanova}.}
  \bibinfo{year}{2019}\natexlab{}.
\newblock \showarticletitle{Latent retrieval for weakly supervised open domain
  question answering}.
\newblock \bibinfo{journal}{\emph{arXiv preprint arXiv:1906.00300}}
  (\bibinfo{year}{2019}).
\newblock


\bibitem[\protect\citeauthoryear{Luan, Eisenstein, Toutanova, and Collins}{Luan
  et~al\mbox{.}}{2020}]%
        {Luan2020SparseDA}
\bibfield{author}{\bibinfo{person}{Yi Luan}, \bibinfo{person}{Jacob
  Eisenstein}, \bibinfo{person}{Kristina Toutanova}, {and} \bibinfo{person}{M.
  Collins}.} \bibinfo{year}{2020}\natexlab{}.
\newblock \showarticletitle{Sparse, Dense, and Attentional Representations for
  Text Retrieval}.
\newblock \bibinfo{journal}{\emph{ArXiv}}  \bibinfo{volume}{abs/2005.00181}
  (\bibinfo{year}{2020}).
\newblock


\bibitem[\protect\citeauthoryear{Nguyen, Rosenberg, Song, Gao, Tiwary,
  Majumder, and Deng}{Nguyen et~al\mbox{.}}{2016}]%
        {NguyenRSGTMD16}
\bibfield{author}{\bibinfo{person}{Tri Nguyen}, \bibinfo{person}{Mir
  Rosenberg}, \bibinfo{person}{Xia Song}, \bibinfo{person}{Jianfeng Gao},
  \bibinfo{person}{Saurabh Tiwary}, \bibinfo{person}{Rangan Majumder}, {and}
  \bibinfo{person}{Li Deng}.} \bibinfo{year}{2016}\natexlab{}.
\newblock \showarticletitle{{MS} {MARCO:} {A} Human Generated MAchine Reading
  COmprehension Dataset} \emph{(\bibinfo{series}{{CEUR} Workshop
  Proceedings})}, Vol.~\bibinfo{volume}{1773}.
  \bibinfo{publisher}{CEUR-WS.org}.
\newblock


\bibitem[\protect\citeauthoryear{Nogueira, Yang, Lin, and Cho}{Nogueira
  et~al\mbox{.}}{2019}]%
        {nogueira2019document}
\bibfield{author}{\bibinfo{person}{Rodrigo Nogueira}, \bibinfo{person}{Wei
  Yang}, \bibinfo{person}{Jimmy Lin}, {and} \bibinfo{person}{Kyunghyun Cho}.}
  \bibinfo{year}{2019}\natexlab{}.
\newblock \showarticletitle{Document Expansion by Query Prediction}.
\newblock \bibinfo{journal}{\emph{arXiv preprint arXiv:1904.08375}}
  (\bibinfo{year}{2019}).
\newblock


\bibitem[\protect\citeauthoryear{Peters, Neumann, Iyyer, Gardner, Clark, Lee,
  and Zettlemoyer}{Peters et~al\mbox{.}}{2018}]%
        {peters2018deep}
\bibfield{author}{\bibinfo{person}{Matthew~E Peters}, \bibinfo{person}{Mark
  Neumann}, \bibinfo{person}{Mohit Iyyer}, \bibinfo{person}{Matt Gardner},
  \bibinfo{person}{Christopher Clark}, \bibinfo{person}{Kenton Lee}, {and}
  \bibinfo{person}{Luke Zettlemoyer}.} \bibinfo{year}{2018}\natexlab{}.
\newblock \showarticletitle{Deep contextualized word representations}.
\newblock \bibinfo{journal}{\emph{arXiv preprint arXiv:1802.05365}}
  (\bibinfo{year}{2018}).
\newblock


\bibitem[\protect\citeauthoryear{Raffel, Shazeer, Roberts, Lee, Narang, Matena,
  Zhou, Li, and Liu}{Raffel et~al\mbox{.}}{2019}]%
        {Raffel2019ExploringTL}
\bibfield{author}{\bibinfo{person}{Colin Raffel}, \bibinfo{person}{Noam
  Shazeer}, \bibinfo{person}{Adam Roberts}, \bibinfo{person}{Katherine Lee},
  \bibinfo{person}{Sharan Narang}, \bibinfo{person}{Michael Matena},
  \bibinfo{person}{Yanqi Zhou}, \bibinfo{person}{W. Li}, {and}
  \bibinfo{person}{Peter~J. Liu}.} \bibinfo{year}{2019}\natexlab{}.
\newblock \showarticletitle{Exploring the Limits of Transfer Learning with a
  Unified Text-to-Text Transformer}.
\newblock \bibinfo{journal}{\emph{ArXiv}}  \bibinfo{volume}{abs/1910.10683}
  (\bibinfo{year}{2019}).
\newblock


\bibitem[\protect\citeauthoryear{Robertson and Walker}{Robertson and
  Walker}{1994}]%
        {robertson1994some}
\bibfield{author}{\bibinfo{person}{Stephen~E Robertson} {and}
  \bibinfo{person}{Steve Walker}.} \bibinfo{year}{1994}\natexlab{}.
\newblock \showarticletitle{Some simple effective approximations to the
  2-poisson model for probabilistic weighted retrieval}. In
  \bibinfo{booktitle}{\emph{SIGIR¡¯94}}. Springer, \bibinfo{pages}{232--241}.
\newblock


\bibitem[\protect\citeauthoryear{Sparck-jones}{Sparck-jones}{1972}]%
        {Sparckjones1972ASI}
\bibfield{author}{\bibinfo{person}{K. Sparck-jones}.}
  \bibinfo{year}{1972}\natexlab{}.
\newblock \showarticletitle{A statistical interpretation of term specificity
  and its application in retrieval}.
\newblock


\bibitem[\protect\citeauthoryear{Strohman, Metzler, Turtle, and Croft}{Strohman
  et~al\mbox{.}}{2005}]%
        {strohman2005indri}
\bibfield{author}{\bibinfo{person}{Trevor Strohman}, \bibinfo{person}{Donald
  Metzler}, \bibinfo{person}{Howard Turtle}, {and} \bibinfo{person}{W~Bruce
  Croft}.} \bibinfo{year}{2005}\natexlab{}.
\newblock \showarticletitle{Indri: A language model-based search engine for
  complex queries}. In \bibinfo{booktitle}{\emph{Proceedings of the
  international conference on intelligent analysis}}, Vol.~\bibinfo{volume}{2}.
  Citeseer, \bibinfo{pages}{2--6}.
\newblock


\bibitem[\protect\citeauthoryear{Zamani, Dehghani, Croft, Learned-Miller, and
  Kamps}{Zamani et~al\mbox{.}}{2018}]%
        {zamani2018neural}
\bibfield{author}{\bibinfo{person}{Hamed Zamani}, \bibinfo{person}{Mostafa
  Dehghani}, \bibinfo{person}{W~Bruce Croft}, \bibinfo{person}{Erik
  Learned-Miller}, {and} \bibinfo{person}{Jaap Kamps}.}
  \bibinfo{year}{2018}\natexlab{}.
\newblock \showarticletitle{From neural re-ranking to neural ranking: Learning
  a sparse representation for inverted indexing}. In
  \bibinfo{booktitle}{\emph{Proceedings of the 27th ACM international
  conference on information and knowledge management}}.
  \bibinfo{pages}{497--506}.
\newblock


\bibitem[\protect\citeauthoryear{Zaragoza, Craswell, Taylor, Saria, and
  Robertson}{Zaragoza et~al\mbox{.}}{2004}]%
        {zaragoza2004microsoft}
\bibfield{author}{\bibinfo{person}{Hugo Zaragoza}, \bibinfo{person}{Nick
  Craswell}, \bibinfo{person}{Michael~J Taylor}, \bibinfo{person}{Suchi Saria},
  {and} \bibinfo{person}{Stephen~E Robertson}.}
  \bibinfo{year}{2004}\natexlab{}.
\newblock \showarticletitle{Microsoft Cambridge at TREC 13: Web and Hard
  Tracks.}. In \bibinfo{booktitle}{\emph{TREC}}, Vol.~\bibinfo{volume}{4}.
  \bibinfo{pages}{1--1}.
\newblock


\end{thebibliography}

\appendix

\end{document}